\documentstyle[epsfig,epsf,ifthen,multicol,aps,prb]{revtex}

\begin{document}

%\draft 

%\twocolumn[\hsize\textwidth\columnwidth\hsize\csname@twocolumnfalse\endcsname

\title{Using Electrons on Liquid Helium for Quantum Computing}

\author{A.J. Dahm$^a$, J.M. Goodkind$^b$, I. Karakurt$^a$, and S. Pilla$^b$}
\address{$^a$Department of Physics, Case Western Reserve University, Cleveland, \\OH 44106-7079,USA\\
$^b$Department of Physics, University of California at San Diego,
La Jolla, \\CA 92093-0319, USA}

\date{November 4, 2001}

\maketitle

\begin{abstract}
We describe a quantum computer based on electrons supported by a helium film and localized laterally by small electrodes just under the helium surface.  Each qubit is made of combinations of the ground and first excited state of an electron trapped in the image potential well at the surface.  Mechanisms for preparing the initial state of the qubit, operations with the qubits, and a proposed readout are described.  This system is, in principle, capable of $10^5$ operations in a decoherence time.

PACS numbers: 03.67.Lx, 73.20.-r.

\end{abstract}

%\vskip[2pc]

\begin{multicols}{2}

\section{INTRODUCTION}

A full description of quantum computing is beyond the scope of this paper.  More complete
descriptions are given elsewhere.\cite{Nielsen}

A classical computer has binary bits with values that are either
$0$ or $1$. A quantum computer is operated with quantum bits,
called qubits.  Each qubit uses two energy levels of a quantum
system for the components $0$ and $1$.  However, a qubit can be in
a state that is a superposition of these two components,
\begin{equation}
\Psi = a |0\rangle + b |1\rangle;  \; \; \; \; \; |a|^2 + |b|^2  = 1
\end{equation}
A quantum computer would use a superposition of many qubit states.
An n-qubit system has $2^n$ basis  vectors, $|x_j\rangle$, which
can be taken as products of the basis vectors of each of the $n$
qubits. An arbitrary combination of these can be written as
\begin{equation}
\Psi = \sum_j \alpha_j |x_j\rangle; \; \; \; \; \; \sum_j|\alpha_j|^{2} = 1
\end{equation}
An example of a basis vector for the five qubit case (for $j = k$)
is, $|x_k\rangle$ = $|01100\rangle$.

The superposition states represented by Eqs. (1) and (2) are
responsible for the great potential advantage of quantum logic
operations relative to classical ones.  In the classical case an
operation starts, proceeds and ends with every qubit in a given
state.  In the quantum case, an operation can use qubits in a
superposition of many possible basis states.  Operation on such
superposition states can be equivalent to performing a large
number of computations in parallel.  The difficulty in utilizing
this advantage arises from the fundamental nature of measurement
in quantum processes, namely that measurement of the energy of an
individual qubit will necessarily collapse the wave function so
that the result can be only either $|0\rangle$ or $|1\rangle$ for
each qubit.  This requires algorithms that can yield definite
answers to computations even though the square of the coefficients
$\alpha_j$ represent only probabilities.

For quantum logic operations with a physical system one must have\\
a) discrete states that can be identified with the components $|0\rangle$ or $|1\rangle$.
\\
b) a method to prepare an initial state.
\\
c) a method for operating quantum gates.
\\
d) a readout mechanism.
\\
e) a coherence time sufficient to undertake a large number of operations.
\\
f) for practical computing the system must be scalable to a large number of qubits.

We describe here a proposed quantum computer with qubits made of
electrons on the surface of a liquid helium film and describe our
method for fabricating the qubits and the methods by which their
quantum states can be manipulated and measured.  This system was
first proposed by Platzman and Dykman\cite{PDScience} and has been
described elsewhere.\cite{PDFort,MJLea,DykmanQI,GoodkindQI}

\section{ELECTRONS ON HELIUM}

\begin{figure}

\centerline{\includegraphics[height=2.1in]{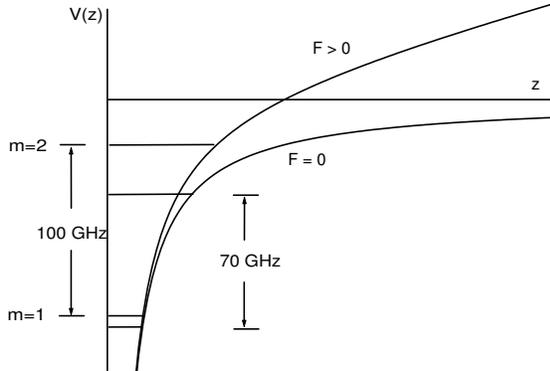}}

%\framebox[5in]{\rule[1.125in]{0in}{1.125in}}

%\makebox[3in]{\rule[1.125in]{0in}{1.125in}}

\caption{The potentials and energy levels with and without an
electric field F applied normal to the surface.  The ground (m=1)
and first excited (m=2) energy levels for each potential are
indicated schematically.}

\label{fig1}

\end{figure}

Electrons are bound to the surface of liquid helium by the
dielectric image potential.  A repulsive Pauli potential prevents
them from penetrating into liquid helium.  The hydrogenic-like
potential is of the form
\begin{equation}
V = -\Lambda e^2/z;  \; \; \; \; \;  \Lambda = (\kappa - 1)/4(\kappa + 1),
\end{equation}
where $z$ is the coordinate normal to the surface, and $\kappa$ is
the dielectric constant of helium.  The energy levels form a
Rydberg spectrum, $E_n = - R/n^2$.  We give parameters for liquid
$^3$He; $\Lambda_3$ = $0.00521$, $R_3$ = $0.37$ meV, and the Bohr
radius for this problem is $a_B$ = $10.2$ nm.  The average
separation of the electron from the surface is $\langle z \rangle$
= $15.3$ nm and $61$ nm for the ground and first excited state,
respectively.  The transition frequency between the ground and
first excited state\cite{Volodin} is $70$ GHz.  These transitions
can be shifted with a Stark field\cite{Grimes} applied normal to
the surface.  The potentials with and without a Stark field are
shown in Fig. \ref{fig1}.

\section{DESIGN OF THE COMPUTER}
We identify the ground and first excited states of these electrons
with the states $|0\rangle$ and $|1\rangle$, respectively.  In
order to address and control the qubits each electron must be
localized laterally.  This will be accomplished by locating
electrons above microelectrodes (posts) that are separated by
about $0.5$ $\mu$m.  The electrons will be separated from the tops
of the posts by a $0.5$ $\mu$m thick helium film.  The lateral
confinement results from the image potentials of the posts and the
electric field from the potential on the posts, and the electron
will be in the ground state for lateral motion at low
temperatures.
\begin{figure}
\centerline{\includegraphics[height=2.3in]{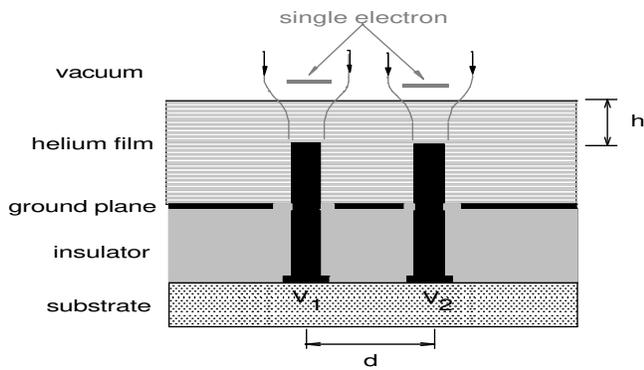}}

%\framebox[5in]{\rule[1.125in]{0in}{1.125in}}

%\makebox[5in]{\rule[1.125in]{0in}{1.125in}}

\caption{The geometry of a two-qubit system with electrons above
the microstructure and the helium film. Electric field lines are
shown.  Drawing is not to scale.  The optimal dimensions are d
$\approx$ h = $0.5$ $\mu$m.  Control potentials $V_1$ and $V_2$
are applied on the micro-electrodes.}

\label{fig2}

\end{figure}

A schematic of posts and electrons for a two-qubit system is shown
in Fig. 2.  A voltage applied to a given post controls the Stark
field for the corresponding electron.  An array of posts
fabricated at the Cornell Nanofabrication Facility is shown in
Fig. 3.  A prototype with lead wires to the posts and an isolated
ground plane to screen the field from the leads has also been
fabricated.  Numerical computations of the electric field from the
posts and ground plane have been completed and fabrication of a
final design is in progress.   Note that this system is scalable
to an arbitrary number of posts or qubits.
\begin{figure}

\centerline{\includegraphics[height=1.6in]{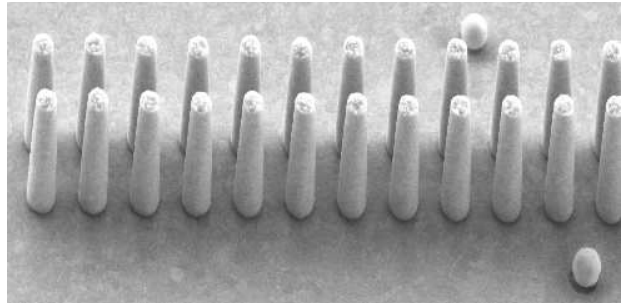}}

%\framebox[5in]{\rule[1.125in]{0in}{1.125in}}

%\makebox[5in]{\rule[1.125in]{0in}{1.125in}}

\caption{SEM image of double row of posts grown on a gold film.
The posts are $1.5$ $\mu$m high and separated by $0.5$ $\mu$m.}

\label{fig3}

\end{figure}

A schematic of our cell is shown in Fig. 4.  The posts will be
located in the bottom of a waveguide that transmits sub-mm
radiation to the electrons.  Superconducting microbolometers will
be located at the top of the guide to detect electrons that are
allowed to escape from the posts.  A tunnel-diode
electron-emission source will be located above the electron
detectors.  Electrons will be loaded onto the film through a hole
in the detector chip, and one electron will be trapped over each
post by an applied positive potential.  The thickness of the
helium film will be measured with a capacitor made of metal strips
deposited on a part of the ground plane containing the posts that
extends outside of the waveguide.  The electrodes of the capacitor
are in the shape of a comb with interwoven teeth spaced by
approximately one micron.  The system will be operated at $10$ mK
to increase coherence times of the qubit states.

Our initial tests will be on a $^3$He film.  The microwave setup
for $^3$He is easier to work with because the transition
frequencies are lower and the corresponding waveguide is larger.
There a number of other advantages in using $^3$He.  Liquid $^3$He
has a large viscosity at low temperatures.  This should severely
dampen high-frequency ripplons involved in T$_1$ and T$_2$
processes, although new decoherence processes involving
interactions with the bulk $^3$He may occur.  Thermal contact
between the sample chamber and $^3$He is much easier to achieve
than for $^4$He, and the surface is less affected by microphonics.
Finally, the larger separation of the electrons from the surface
leads to a somewhat larger dipole interaction between neighboring
states.
\begin{figure}

\centerline{\includegraphics[height=1.8in]{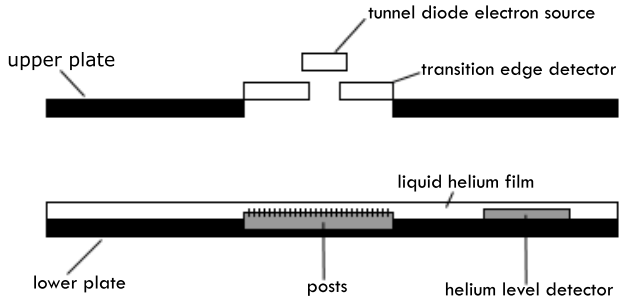}}

%\framebox[5in]{\rule[1.125in]{0in}{1.125in}}

%\makebox[5in]{\rule[1.125in]{0in}{1.125in}}

\caption{Schematic of the cell. The upper plate includes detectors used in the readout. The lower plate
includes the posts and is covered with the helium film.  The electrons float over the posts about $10$ nm
above the surface of the helium film.}

\label{fig4}

\end{figure}

\section{OPERATIONS}

The operation would normally begin with all qubits in the ground
state.  Then according to the requirements of the desired
operations, each qubit would be prepared in some admixture of
states $|0\rangle$ and $|1\rangle$ by Stark shifting individual
qubits into resonance with microwave radiation for a predetermined
time.\cite{PDFort,DykmanQI}  The state of the qubit $n$ will be
\begin{equation}
\Psi_n =  cos(\Theta_n/2)|0\rangle - i\; sin(\Theta_n/2)|1\rangle,
\end{equation}
where $\Theta_n = \Omega\tau_n$, $\Omega = eE_{rf}\langle1|z|2\rangle/\hbar$ is the Rabi frequency, $E_{rf}$ is the strength of the rf field, and $\tau_n$ is the time the $n^{th}$ qubit is in resonance with the microwave field.

In general, computations will be implemented by applying pulses of
radiation to interacting qubits.  We illustrate a potential
operating mode of the system by describing two qubits operated as
a swap gate.  The interaction between qubits is via the Coulomb
energy, which is much larger than the interaction between the
induced dipole moment of each qubit.  The dipolar component of the
direct interaction potential between qubits $i$ and $j$ is
\begin{equation}
V(z_i, z_j) \sim (e^2/8\pi \varepsilon_0 d^3)(z_i-z_j)^2,
\end{equation}
where $d$ is the electron separation, and $z_i$ is the separation
of the $i^{th}$ electron from the helium surface.  Start with one
qubit in the state  $|0\rangle$ and the other in the state
$|1\rangle$.  Next apply the same Stark fields to both qubits so
that the states $|01\rangle$ and $|10\rangle$ would be degenerate.
In this condition the system will oscillate between the two states
at a frequency given by the interaction energy, which in first
order is given by $\sim e^2a_B^2/4\pi\varepsilon_0d^3$. This
frequency is $\sim 1$ GHz for a separation of $0.5$ $\mu$m. By
leaving the electric fields in this condition for one half cycle
of this oscillation, the two qubits will swap states.  It will be
difficult to tune neighboring qubits to precisely identical Stark
shifts, and in practice we may sweep the Stark shift of one qubit
through resonance with a neighboring qubit.\cite{DykmanQI}  In
this case, the final state of the qubit will depend on the rate at
which the electric field is swept through the resonance condition.

Readout.  The wavefunction of the system of entangled qubits
collapses when a measurement is made.  Thus, the states of all
qubits must be read within the time scale is set by the plasma
frequency $\sim 100$ GHz.  We describe here our initial proposal
for a destructive readout pending research into other schemes.  We
will apply a short, $\sim 1$ ns, ramp of an extracting electric
field to all qubits.  The potential for a fixed value of
extracting field is shown in Fig. 5.  The tunneling probability is
exponential in the time-dependent barrier height.  All electrons
in the upper $|1\rangle$ state will tunnel through the barrier
within a short period of time when this probability becomes
sufficiently large.  The escape of an electron will be detected by
the bolometer detector mentioned above.  For this extracting field
the tunneling probability will be negligibly small for electrons
in the ground $|0\rangle$ state.  After the ramp is removed the
remaining electrons will be in the ground state.  Subsequently, we
plan to sequentially apply to each post an extracting field
sufficiently large so that electrons in the ground state will
tunnel from the surface.\cite{Saville}  A $|0\rangle$ will be
registered for each electron detected and a $|1\rangle$ for those
states that are empty.

An alternate scheme would be to apply an extracting voltage to the
post under one electron at a time.  For small numbers of qubits in
our initial exploratory experiments, the system would then be
prepared in the same state repeatedly and a different qubit
sampled each time the operation is carried out.  Ultimately we
hope to develop a non-destructive readout that will allow
simultaneous measurement of the states of all qubits without
allowing the electrons to escape.
\begin{figure}

\centerline{\includegraphics[height=2.1in]{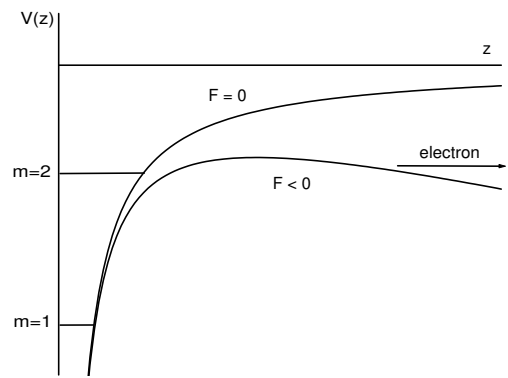}}

%\framebox[5in]{\rule[1.125in]{0in}{1.125in}}

%\makebox[5in]{\rule[0.125in]{2in}{0.125in}}

\caption{The potential with an extracting field.}

\label{fig5}

\end{figure}

\section{DECOHERENCE}
All logic operations must be accomplished in less time than it
takes for the interactions of the qubit with the environment to
destroy the phase coherence of the state functions.  For electrons
on $^4$He the lifetime of the excited state $|1\rangle$ is limited by
interactions with ripplons.  The electron-ripplon coupling
Hamiltonian\cite{PDScience,PDFort,DykmanQI} is
\begin{equation}
H_{er} = eE_\perp\delta,
\end{equation}
where $E_\perp$ is the normal component of the electric field that
includes both the applied field and variations in the helium
dielectric image field due to surface distortions, and $\delta$ is the
amplitude of the surface height variation.  The average rms
thermal fluctuation of the surface is
\begin{equation}
\delta_\textnormal{T} = (k_B\textnormal{T}/\sigma)^{1/2} \cong 2\times10^{-9} cm
\end{equation}
The transition from the excited to ground state requires a ripplon
with a wave vector $\cong a_B^{-1}$.  For a single electron on
bulk helium a radiationless transition occurs with the energy
absorbed by electron plane-wave states for motion parallel to the
surface and momentum absorbed by ripplons.  For this case a
calculation of T$_1$ yields
\begin{equation}
\textnormal{T}_1^{-1} \cong \Delta\nu(\delta_T/a_B)^2,
\end{equation}
where $\Delta\nu$ is the transition frequency.  At T$ = 10$ mK,
$\delta_T/a_B \cong 10^{-3}$, and T$_1$ $\cong 10 \mu$s.

For electrons confined by posts, the lateral states are harmonic
oscillator states of the image potential well of the posts.  These
are separated in energy\cite{Dykman2} by $\hbar\Omega_1 \approx
\hbar(e^2/4 \pi \varepsilon_0 md^3)^{1/2} \sim 300$ mK.  For an
array of electrons there is a band of plasma oscillations
associated with each harmonic oscillator level, which for an
electron separation of $0.5$ $\mu$m on bulk helium has a band
width that is greater than $300$ mK.  The confining potential of
the posts reduces the width of this band.  Preliminary
calculations\cite{Dykman2} suggest that the width may be
sufficiently small to prevent conservation of energy in a
radiationless transition.  This is illustrated in Fig. 6 where the
transition is shown to be incommensurate in energy with an
excitation between the harmonic oscillator states. Suppression of
this relaxation channel would lead to a very large value of T$_1$.
\begin{figure}

\centerline{\includegraphics[height=2.0in]{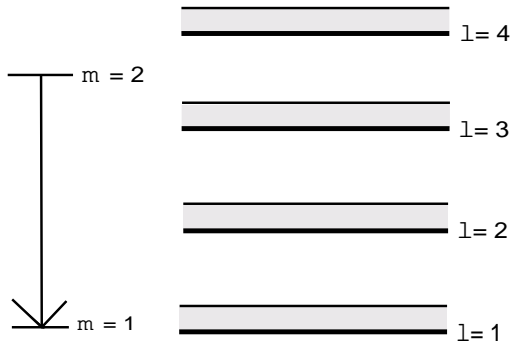}}

%\framebox[5in]{\rule[1.125in]{0in}{1.125in}}

%\makebox[5in]{\rule[0.125in]{2in}{0.125in}}

\caption{Energy level schematic.  The integer m labels the hydrogenic-like states, and the integer l labels lateral states, which are harmonic oscillator states in the image field of the posts or Landau levels in a magnetic field.  The hatched region indicates the band of plasma oscillations associated with each level.}

\label{fig6}

\end{figure}

If the confining potential does not sufficiently reduce the width
of the plasma band, this will be accomplished with an applied
magnetic field.  A magnetic field confines the lateral states to
Landau levels with a band of plasma oscillations of width
$\hbar\Omega_2 \sim h\omega_p^2/\omega_c$.  Here $\omega_p$ and
$\omega_c$ are, respectively, the zone-boundary plasmon of the
ordered qubit array and cyclotron frequencies.  The transitions
between Landau levels with the emission of one ripplon cannot
conserve energy and momentum for $B \cong 1.5$ Tesla.

The two-ripplon process can either cause a transition to the
ground state or lead to dephasing by an incoherent phase
modulation due to quasi-elastic scattering within a "hydrogenic
level" by thermal ripplons with a wave vector equal to the inverse
cyclotron radius\cite{DykmanS}.  At $10$ mK and a field of $1.5$
Tesla, this leads to a dephasing time\cite{Dykman2} T$_2$ of $\sim
100$ ms.  Single operations can be made in $1$ ns.  Thus, in
principle, $\sim 10^5$ operations can be preformed in a
decoherence time.

High frequency ripplons are strongly damped on a liquid $^3$He
surface.  Thus, it is possible that the values of T$_1$ and T$_2$ may
be longer than for $^4$He.  A possible dephasing mechanism for the
case of $^3$He is via interactions of the electron with
excitations on the Fermi surface of bulk $^3$He.  Theoretical
calculations of this mechanism have not been carried out.

\section*{ACKNOWLEDGMENTS}

The authors wish to acknowledge Mark Dykman for helpful
conversations.  This work was supported in part by NSF grant
EIS-0085922.

\noindent
\vspace{.5 cm}

\bibliographystyle{prsty}

\end{multicols}

\end{document}